%% file: lymanalphaz7.tex
\documentclass[twocolumn, times, tighten]{aastex63}

\usepackage{graphicx}	
\usepackage{amsmath}	
\usepackage{amssymb}	
\usepackage{footnote}
\usepackage{tablefootnote}
\usepackage{booktabs}

\newcommand{\kms}{km\,s$^{-1}$}

\received{}
\revised{}
\accepted{}
\submitjournal{ApJL}

\shorttitle{Dichromatic Primeval Galaxy at $z\sim 7$}
\shortauthors{Debora Pelliccia et al.}

\begin{document}

\title[RELICS-DP7]{RELICS-DP7: Spectroscopic Confirmation of a Dichromatic Primeval Galaxy at $z\sim 7$}

\author[0000-0002-3007-0013]{Debora Pelliccia}\email{dpelliccia@ucolick.org}
\affiliation{UCO/Lick Observatory, Department of Astronomy \& Astrophysics, UC Santa Cruz, 1156 High Street, Santa Cruz, CA, 95064, USA}\affiliation{Department of Physics \& Astronomy, University of California, Davis, One Shields Ave, Davis, CA 95616, USA}

\author[0000-0002-6338-7295]{Victoria Strait}\email{vbstrait@ucdavis.edu}\affiliation{Department of Physics \& Astronomy, University of California, Davis, One Shields Ave, Davis, CA 95616, USA}

\author[0000-0002-1428-7036]{Brian C. Lemaux}\affiliation{Department of Physics \& Astronomy, University of California, Davis, One Shields Ave, Davis, CA 95616, USA}

\author[0000-0001-5984-0395]{Maru{\v s}a Brada{\v c}}\affiliation{Department of Physics \& Astronomy, University of California, Davis, One Shields Ave, Davis, CA 95616, USA}

\author[0000-0001-7410-7669]{Dan Coe}\affiliation{Space Telescope Science Institute, 3700 San Martin Drive, Baltimore, MD 21218, USA}

\author[0000-0002-7365-4131]{Patricia Bolan}\affiliation{Department of Physics \& Astronomy, University of California, Davis, One Shields Ave, Davis, CA 95616, USA}

\author[0000-0002-7908-9284]{Larry D. Bradley}\affiliation{Space Telescope Science Institute, 3700 San Martin Drive, Baltimore, MD 21218, USA}

\author[0000-0003-1625-8009]{Brenda Frye}\affiliation{Department of Astronomy, Steward Observatory, University of Arizona, 933 North Cherry Avenue, Tucson, AZ, 85721, USA}

\author[0000-0003-0965-605X]{Pratik J. Gandhi}\affiliation{Department of Physics \& Astronomy, University of California, Davis, One Shields Ave, Davis, CA 95616, USA}

\author{Ramesh Mainali}\affiliation{Observational Cosmology Lab, NASA Goddard Space Flight Center, 8800 Greenbelt Rd.,Greenbelt, MD 20771, USA}

\author[0000-0002-3407-1785]{Charlotte Mason}\affiliation{Harvard-Smithsonian Center for Astrophysics, 60 Garden
St, Cambridge, MA, 02138, USA}\affiliation{Hubble Fellow}

\author{Masami Ouchi}\affiliation{Institute for Cosmic Ray Research, The University of Tokyo, 5-1-5 Kashiwanoha, Kashiwa, Chiba 277-8582, Japan}\affiliation{Kavli Institute for the Physics and Mathematics of the Universe (Kavli IPMU, WPI), University of Tokyo, Kashiwa, Chiba 277-8583, Japan}

\author[0000-0002-7559-0864]{Keren Sharon}\affiliation{Department of Astronomy, University of Michigan, 1085 S. University Ave, Ann Arbor, MI 48109, USA}

\author{Michele Trenti}\affiliation{School of Physics, University of Melbourne, VIC 3010, Australia}\affiliation{ARC Centre of Excellence for All Sky Astrophysics in 3 Dimensions (ASTRO 3D), VIC 2010, Australia}

\author[0000-0002-0350-4488]{Adi Zitrin}\affiliation{Department of Physics, Ben-Gurion University, Beer-Sheva 84105, Israel}



\begin{abstract}
We report the discovery of a spectroscopically-confirmed strong Lyman-$\alpha$ emitter at $z=7.0281\pm0.0003$, observed as part of the Reionization Cluster Lensing Survey (RELICS). This galaxy, dubbed ``Dichromatic Primeval Galaxy" at $z\sim7$ (DP7), shows two distinct components. While fairly unremarkable in terms of its ultraviolet (UV) luminosity ($\sim0.3L^{\ast}_{UV}$, where $L^{\ast}_{UV}$ is the characteristic luminosity), DP7 has one of the highest observed Lyman-$\alpha$ equivalent widths (EWs) among Lyman-$\alpha$ emitters at $z>6$ ($>200$\AA\ in the rest frame). The strong Lyman-$\alpha$ emission generally suggests a young metal-poor, low-dust galaxy; however, we find that the UV slope $\beta$ of the galaxy as a whole is redder than typical star-forming galaxies at these redshifts, $-1.13\pm 0.84$, likely indicating, on average, a considerable amount of dust obscuration, or an older stellar population. When we measure $\beta$ for the two components separately, however, we find evidence of differing UV colors, suggesting two separate stellar populations. Also, we find that Lyman-$\alpha$ is spatially extended and likely larger than the galaxy size, hinting to the possible existence of a Lyman-$\alpha$ halo. Rejuvenation or merging events could explain these results. Either scenario requires an extreme stellar population, possibly including a component of Population III stars, or an obscured Active Galactic Nucleus. DP7, with its low UV luminosity and high Lyman-$\alpha$ EW, represents the typical galaxies that are thought to be the major contribution to the reionization of the Universe, and for this reason DP7 is an excellent target for follow-up with the \textit{James Webb Space Telescope}.

\end{abstract}

\keywords{Galaxies, High-redshift galaxies, Galaxy evolution, Reionization}


\section{Introduction} \label{sec:intro}
Spectroscopic observations of distant galaxies ($z>6$) allow us to constrain early galaxy formation and the epoch of reionization. In particular, the $\lambda$1215.7\AA\ Lyman-$\alpha$ line is both the strongest recombination line of hydrogen intrinsically and resonantly scattered by neutral hydrogen, making its strength a sensitive gauge of the amount of neutral hydrogen in this epoch.  The study of Lyman-$\alpha$ emission in galaxies is commonly used to constrain the neutral fraction of the InterGalactic Medium (IGM) and, hence, different reionization timelines \citep[e.g.,][]{robertson15,Mason2019}. Identifying ionized and neutral regions of the IGM not only characterizes the topology of reionization but also allows for the identification of the type of sources that likely drove reionization and the properties of their corresponding ionized bubbles \citep[e.g.,][]{malhotra06,Mason2020}.

While Lyman-$\alpha$ is the most commonly detected spectral line for galaxies spectroscopically confirmed at $z>7$, confirmation based on this line has still
proven difficult at these redshifts
\citep[e.g.,][]{stark10,pentericci14,schenker14,hoag2019, jung20}. The reason is likely in part due to Lyman-$\alpha$ being scattered by a patchy neutral medium before reionization was complete \citep[e.g.,][]{treu13,mason18a}, in addition to the effect of increasing number, strength and variability of OH lines around the expected Lyman-$\alpha$ wavelength at these redshits. The exceptions are typically galaxies which have carved out large enough ionized bubbles for Lyman-$\alpha$ emission to escape, allowing it to be observed \citep[e.g.,][]{Finkelstein2013, Zitrin2015, Oesch2015, Roberts-Borsani2016}.
In addition to adding constraining power to the timing of the epoch of reionization (e.g., \citealp{Mason2019,hoag2019, jung20}), some recent studies have used measurements of Lyman-$\alpha$ strengths and spatial extent to estimate the sizes of these ionized bubbles (e.g., \citealp{Tilvi2020}), measure ionization parameters (e.g., \citealp{Matthee2017b}), identify extremely young, metal-poor stellar populations (e.g., \citealp{Ouchi2013,Sobral2015,Matthee2019,Matthee2020}), and identify possible candidates that contain Population III (PopIII) stars (e.g., \citealp{Sobral2015,Vanzella2020}). With spectrally-resolved Lyman-$\alpha$ emission, some studies have shown that it is possible to measure the residual neutral fraction within the bubble, the bubble size, as well as physical conditions in galaxies that incite the formation of such bubbles (e.g., \citealp{Verhamme2015,Mason2020}).

A large fraction of $z\sim7$ galaxies with extreme Lyman-$\alpha$ emission consist of multiple components in the rest-frame UV \citep[e.g,][]{Ouchi2013, Sobral2015}, which makes it difficult to interpret spatially unresolved Lyman-$\alpha$ emission. However, the high incidence of such multi-component systems suggests that mergers or accretion events \citep[e.g,][]{Ouchi2013, Matthee2020}, which are believed to cause an increase or rejuvenation in star formation, are perhaps a common, if not necessary, condition for such extreme Lyman-$\alpha$ emission. Additionally, most of these galaxies are comparatively more luminous than the characteristic galaxy for their redshift ($L>L_{UV}^*$, where $L_{UV}^*$ is the characteristic luminosity, see, e.g., \citealt{Matthee2020}), which may be a consequence of clustering effects or an effect of selection.

In this Letter, we present the Dichromatic Primeval Galaxy at $z\sim7$ (DP7), a UV-faint galaxy (0.3$L/L_{UV}^*$) with multiple components, detected with extreme Lyman-$\alpha$ emission from Keck. In Section~\ref{sec:data} we describe available data, in Section~\ref{sec:analysis} we show the resulting spectrum and analysis of the stellar population in DP7 and we conclude in Section~\ref{sec:conclusion}. Throughout the manuscript, we adopt a \citet{Chabrier2003} initial mass function and a $\Lambda$CDM cosmology with \mbox{H$_0$ = 70 km s$^{-1}$}, $\Omega_\Lambda$ = 0.7, and $\mathrm{\Omega_M}$ = 0.3. Magnitudes are given in the AB system. Distances are quoted as proper distances.

\section{Data} \label{sec:data}
This study is based on data taken from the Reionization Lensing Cluster Survey \citep[RELICS,][]{Coe2019}.  RELICS is a 188-orbit \textit{Hubble Space Telescope} (\textit{HST}) Treasury Program targeting 41 massive galaxy clusters at 0.182$\leq z \leq$0.972. 
RELICS clusters were observed  with the Advanced Camera for Surveys (ACS) and the infrared Wide Field Camera 3 (WFC3/IR) spanning $0.4-1.7\mu m$. Details on the survey can be found in \citet{Coe2019}. Every RELICS cluster was also observed with \textit{Spitzer}/IRAC by a combination of RELICS programs (PI Soifer, PI Brada\v{c}) and archival ones. 

The primary aim of RELICS is to systematically search for lensed high-redshift galaxies in the epoch of reionization. More than 300 candidate galaxies have being discovered with photometric redshifts $z_{phot}>5.5$, using \textit{HST} and \textit{Spitzer} data \citep[see][for details about the selection criteria]{Salmon2020}. In this work we focus on one of the high-redshift candidates form the \citeauthor{Salmon2020} catalog, namely MS1008-12-427, which we dubbed RELICS-DP7 (i.e., RELICS Dichromatic Primeval galaxy at $z\sim7$).  This object is located behind the $z=0.306$ MS~1008.1-1224 cluster. We followed this object up with additional \textit{Spitzer}/IRAC data and Keck spectroscopy, which allowed us to better characterize the properties of this galaxy at the epoch of reionization.

\subsection{HST and Spitzer Data and Photometry} \label{subsec:imaging}
We made use of the \textit{HST} reduced images publicly available via the Mikulski Archive for Space Telescopes (MAST\footnote{https://archive.stsci.edu/prepds/relics/}). 
These were obtained combining all the RELICS and archival ACS (F435W, F606W, F814W) and WFC3/IR (F105W, F125W, F140W, F160W) images available in the MS~1008.1-1224 field of view. Details on the data reduction can be found in \citet{Coe2019}. While DP7 was selected for spectroscopy follow-up by using the original \citeauthor{Salmon2020} catalog, we have reanalyzed the imaging data for this study.

\begin{figure}
\centering
	\includegraphics[width=\linewidth]{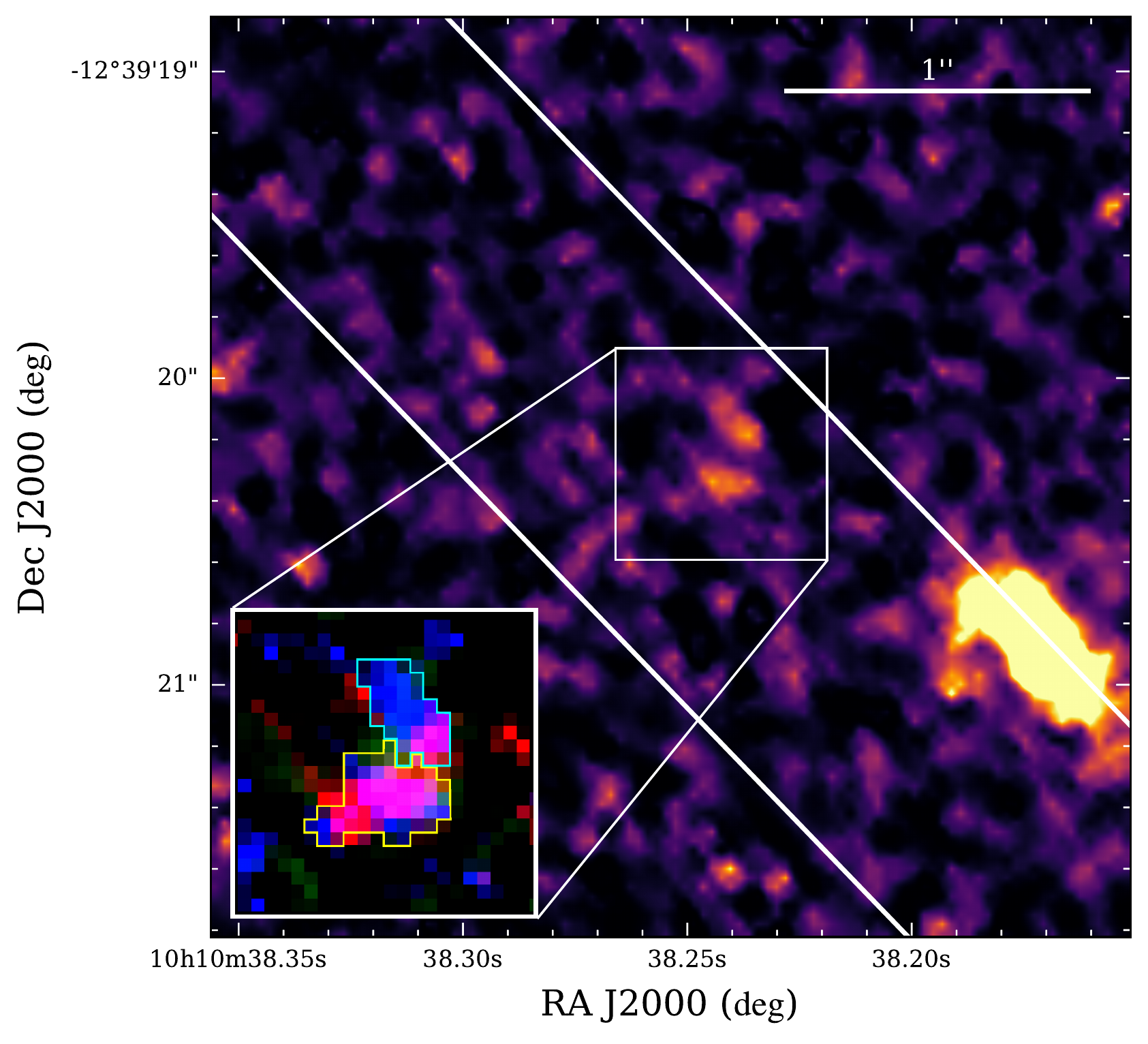}
    \caption{\textit{HST} F105W postage stamp (3\arcsec$\times$3\arcsec) centered on DP7 with superimposed in white the LRIS slit. \textbf{Inset:} \textit{HST} color (F105W + F140W + F160W) postage stamp (0.7\arcsec$\times$0.7\arcsec) showing a southern red component and a northern blue component with superimposed segmentation maps from {\tt SExtractor} in yellow and cyan, respectively.}
    \label{fig:hstimage}
\end{figure}

We run {\tt SExtractor} on the drizzled images with 0.03\arcsec\ resolution to produce a photometric catalogs of our galaxy. In addition, we re-run {\tt SExtractor} to deblend the two components that we can clearly see in the F105W image (see Figure~\ref{fig:hstimage}), by setting the deblending parameters to the following: {\tt DEBLEND\_MINCONT = 0.00005, DEBLEND\_NTHRESH=3}. The inset of Figure~\ref{fig:hstimage} shows the segmentation maps of a northern component (in cyan) and a southern component (in yellow). The WFC3/IR photometry measurements for the two individual components, along with the measurements for the entire galaxy, are reported in Table~\ref{tab:photometry} and were used to investigate the stellar properties of DP7 (see Section~\ref{subsec:betaslope}).
\input{DP7_mags_no_mucorr} \label{tab:photometry}

To derive realistic uncertainties on the fluxes, we accounted for the correlated pixel noise by re-scaling the RMS map that we input in {\tt SExtractor} following the procedure described by \citet{Trenti2011}. The re-scaling factor is computed for each \textit{HST} image, and it is such that the median error quoted by {\tt SExtractor} for the photometry in empty sky apertures, of size comparable to our galaxy, is equal to the RMS of the measured flux in the same aperture.

Following the methodology described in \citet{Fuller2020} (see their Eq.4), we used the F160W apparent magnitude and the $k$-correction adopted by \citeauthor{Fuller2020} to estimate the UV absolute magnitude $M_{UV}$.  This resulted in an $M_{UV}=-19.5\pm0.2\,$mag for the whole galaxy, corresponding to 0.26$L_{UV}^*$, where $L^*$ is the characteristic UV luminosity of a typical galaxy at $z=7$ from \citet{Bouwens2015a}. Since DP7 appears to be formed by two components, from which we were able to measure the photometry separately (see above), we estimated $M_{UV}$ to be equal to $-18.22\pm0.42\,$mag for the northern component, and $-19.1\pm0.22\,$mag for the southern component, corresponding to 0.08 and 0.18$L_{UV}^*$, respectively. All the reported values are corrected for magnification (see Section~\ref{subsec:SED_lensing} and Table~\ref{tab:properties}).

\emph{Spitzer}/Infrared Array Camera (IRAC) images come from a combination of S-RELICS (PI Brada{\v c}, \#12005, \#14017) and Director's Discretionary Time (PI Soifer, \#12123). The MS~1008.1-1224 cluster reached a total of 28 hours of exposure time in each IRAC channel (3.6$\mu$m and 4.5$\mu$m, [3.6] and [4.5]). To reduce and mosaic Spitzer  images, we closely follow the process described by \citet{Bradac2014}.  We create the mosaic images using the MOsaicker and Point source EXtractor (MOPEX)  command-line tools and largely follow the process described in the IRAC Cookbook\footnote{http://irsa.ipac.caltech.edu/data/SPITZER/docs/\\dataanalysistools/cookbook/} for the COSMOS medium-deep data. 

We extract \emph{Spitzer} fluxes following \cite{Strait2020b}. Briefly, we use T-PHOT which was designed to use a high resolution image (in our case, \textit{HST} F160W image and WFC3IR total segmentation) as a prior for reconstructing a model of a low resolution image (in our case, IRAC [3.6] and [4.5]). For field images, T-PHOT is normally run on an entire image at once. However, because background and intracluster light (ICL) varies in a cluster environment, we run T-PHOT individually for each object on a small FOV of 20\arcsec.  T-PHOT requires a PSF of the low resolution image in order to be convolved with the high resolution image. To create a \emph{Spitzer} PSF for this field, we stack point sources from the field, identified with the stellar locus of a flux radius vs. magnitude plot. We require that there are at least 40 point sources in the making of the PSF \citep[see][for more detail]{Strait2020b}.

The two components of DP7 are too close to separate in \emph{Spitzer}/IRAC, so we only report the photometry of the whole galaxy (see Table~\ref{tab:photometry}). The color of DP7 in the IRAC bands ($\rm{[3.6]-[4.5] = 0.1\pm0.5\,mag}$) is consistent with the predicted IRAC color of young star-forming galaxies at $z\sim7$ as well as those of many other galaxies or candidates measured at these redshifts (see \citealp{Strait2020b} and references therein). 

\begin{figure*}
\centering
	\includegraphics[width=\textwidth]{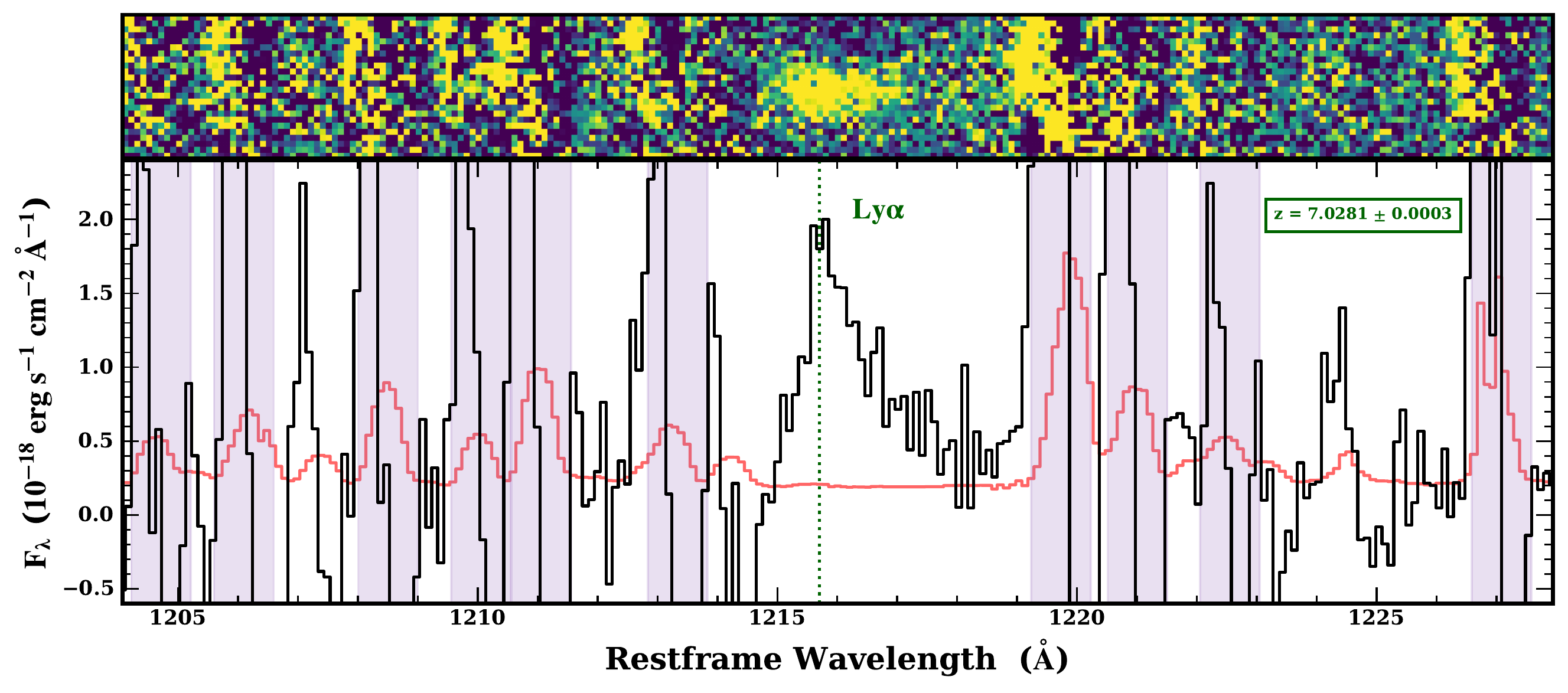}
    \caption{DP7's 2D (top) and 1D (bottom) spectrum cutout centered at Lyman-$\alpha$. The black and red lines show the flux and the noise rest-frame spectrum, respectively. The shaded regions identify the location of the sky lines.}
    \label{fig:spec}
\end{figure*}

\subsection{Spectroscopy} \label{subsec:spectroscopy}
Spectroscopic follow up of DP7 was carried out with the Low Resolution Imaging Spectrometer (LRIS) on Keck~I telescope. Multi-object observations were obtained using 1\arcsec~wide slits over the course of three nights (April 7-9, 2019) for a total of seven hours of integration time (21 frames of 1200s). We used the d680 dichroic, the 300/5000 grism on the blue side covering a range of $\sim4000-7000\,$\AA, and the 600/10000 grating on the red side with a typical wavelength coverage of $\sim7000-10000\,$\AA. Because high-redshift galaxies show spectral features mostly at longer wavelengths (e.g., Lyman-$\alpha$ at $z\geq6$ has $\lambda_{obs}\geq8500\,$\AA), we focus here on the data obtained by the red arm of the spectrograph, which provided 2D spectra with an image scale of 0.135\arcsec pixel$^{-1}$, a spectral scale of 0.8\AA\,pixel$^{-1}$, and full width at half-maximum (FWHM) spectral resolution of $\sim$4.7\AA.

Data reduction was performed using the newly developed open source Python Spectroscopic Data Reduction Pipeline \citep[\texttt{PypeIt};][]{Prochaska2020}. \texttt{PypeIt} applies standard reduction techniques to each observed frame: slit's edges tracing, wavelength calibration using arc frames, flat-field correction, and sky subtraction. These steps produce a 2D and 1D spectrum for each frame. The 1D spectra are extracted using the optimal spectrum extraction technique, and then flux calibrated using the standard star Feige~34 observed during the same nights of our science frames. Finally, we stacked all the flux calibrated 1D spectra, as well as the 2D spectra. 

We estimated the seeing integrated over the entire exposure time, by identifying a star in our observed mask and determined the FWHM, by fitting a Gaussian function to its spatial profile. We selected our star by searching amongst the most likely point sources from the Probabilistic Classifications of Unresolved Point Sources in PanSTARRS1 \citep{TachibanaMiller2018}. We found that the seeing amounted to $0.92\arcsec\pm0.01\arcsec$. We chose not to apply a correction for slit loss due to the possible difference in UV vs. Lyman-$\alpha$ size (see Section~\ref{subsec:lya_size}). However, it is likely to be fairly small ($\sim$15\%, see \citealt{Lemaux2009}), and any slit loss correction would only serve to increase the estimated line strength and would not meaningfully affect our results.

Inspecting our spectra we found that DP7 shows a prominent Lyman-$\alpha$ emission at $\lambda = 9759.5\pm0.4\,$\AA, placing this galaxy at $z=7.0281\pm0.0003$. We determined the redshift by using a $\chi^2$ fitting technique that employed empirical models of lower-redshift ($z\sim1$) emission line galaxies and high-resolution empirical Lyman-$\alpha$ templates from \citet{Lemaux2009}. 
The spectrum was fit over the wavelength range 9200\AA\ to 10000\AA\ and the redshift was allowed to float over the range $6.5 \le z \le 7.1$. Note that limiting the redshift range such that the observed line was forced to be the [OII] $\lambda$3726,3729\AA\ doublet resulted in a worse $\chi^2$ (2.95 vs 1.10 for [OII] and Lyman-$\alpha$, respectively). As an additional check, we estimated the asymmetry of the line profile following the prescription described in \citet{Lemaux2009} and found a value of $1/a_{\lambda}=0.28$, where $a_{\lambda}$ is the asymmetry parameter. The above tests strongly indicate that the identity of the observed line is Lyman-$\alpha$. We also inspected the spectrum obtained with LRIS blue arm and we did not find any significant emission at or near the spatial location of Lyman-$\alpha$, which excludes the possibility of foreground interlopers.
Figure \ref{fig:spec} shows a cutout of the final reduced 2D and 1D spectrum for DP7 centered at Lyman-$\alpha$. 

From Figure~\ref{fig:hstimage}, which shows the position of the slit on the F105W image, we can see a brighter galaxy to the SW, just at the edge of the slit. Our inspection of the 2D spectrum reveled that this galaxy shows, indeed, a faint but clear H$\beta$+[OIII] emission at $z=0.79$. This emission appears to be at 1.01\arcsec$\pm$0.08\arcsec~from the Lyman-$\alpha$ emission, which is consistent with the expected distance from the \textit{HST} image, after taking into account an additional uncertainty of $\sim$0.43\arcsec\ (approximate major axis of the bright galaxy) on the exact location of the emission within the galaxy.

\section{Analysis} \label{sec:analysis}
We describe here the analysis performed to investigate the nature of this high redshift galaxy. A summary of all DP7 properties is reported in Table~\ref{tab:properties}.
\subsection{SED Fitting and Lensing Model} \label{subsec:SED_lensing}
To calculate stellar properties of the whole galaxy, we use a set of $\sim$2000 BC03 stellar population synthesis templates with emission lines. We assume a Chabrier initial mass function between 0.1 and 100 $M_\odot$, a metallicity of 0.2$\,Z_{\odot}$, a constant star formation history, a Small Magellenic Cloud dust law with E$_*$(B-V)=$\mathrm{E_{gas}}$(E(B-V)) with step sizes of $\Delta$E(B-V)=0.05 for 0-0.5 mag and 0.1 for 0.5-1 mag. We allow age to range from 10 Myr to the age of the universe at the redshift of the source ($\sim$750 Myr).  Because nebular emission and continuum can have a non-trivial effect on broadband fluxes \citep[e.g.,][]{Smit14}, we add them by first calculating the hydrogen recombination line strength following the relation from \cite{Leitherer95}, scaling from integrated Lyman-continuum flux, and then following the strengths determined with nebular line ratios by \cite{Anders03}. In addition to nebular emission, we add Lyman-$\alpha$ to the templates; we calculate expected strengths using the ratio with H$\alpha$ and assuming a Case B recombination with a Lyman-$\alpha$ escape fraction of 0.2. While the Lyman-$\alpha$ escape fraction is likely to be in excess of this value for DP7 given the large Lyman-$\alpha$ equivalent width \citep[see Section~\ref{subsec:lya_ew} and the relation of][]{Sobral_Matthee2019}, this assumption enters only for our SED fitting, and our SED-fit results do not change meaningfully if a larger value is chosen.

To correct for the effects of gravitational lensing on the object, we use a lens model of the cluster MS~1008.1-1224 provided by the RELICS team for public use on MAST\footnote{https://archive.stsci.edu/prepds/relics/}. This lens model was created using {\tt Lenstool}. The process closely follows that of \citet{Cerny18} and \citet{Sharon20}. The model uses as constraints four families of multiply-imaged lensed galaxies, two of which are spectroscopically confirmed. We estimated that DP7 is affected by a magnification $\mu=1.15 \pm 0.02$. The statistical uncertainties on $\mu$ are estimated from the magnifications derived from 100 lens models that were sampled from the MCMC chain; the 68\% confidence limits are quoted.

\input{properties} \label{tab:properties}
\subsection{Ly\texorpdfstring{$\alpha$}{alpha} Flux and Equivalent Width} \label{subsec:lya_ew}
We estimated Lyman-$\alpha$ line flux by inspecting the 1D spectrum and selecting a number of bandpasses. We selected a ``feature" bandpass defined to include the spectral line, and four ``continuum" bandpasses, two blueward and  two redward of the emission line, which are used to estimate the background (since the spectrum doesn't show any stellar continuum emission) across the spectral feature. The ``continuum" bandpasses were chosen to be free of sky lines and as close to the emission line in the wavelength direction as the data would allow. We perform a polynomial (order=1) fit to the ``continuum" regions to estimate the background across the emission line, and subtracted it from the flux in the ``feature" bandpass. The total Lyman-$\alpha$ (Ly$\alpha$) line flux was measured as:
\begin{equation}
    \mathrm{F_{Ly\alpha}}= \frac{1}{\mu}\sum_{i=0}^{n}(f_{\lambda,i} - b_{\lambda,i})\delta \lambda_i 
\end{equation}
where $\mu$ is the magnification (Section~\ref{subsec:SED_lensing}), $f_{\lambda,i}$, $b_{\lambda,i}$, and $\delta \lambda_i$ are  the values of flux, background level and pixel scale (\AA\ pixel$^{-1}$), respectively, in the $i$th spectral pixel in the ``feature" bandpass. We estimated $\mathrm{F_{Ly\alpha}}$ to be equal to $\mathrm{1.74\pm 0.17 \times 10^{-17}erg\,s^{-1}\,cm^{-2}}$, which corresponds to a luminosity $\mathrm{L_{Ly\alpha}=1.0\pm 0.1 \times 10^{43}erg\,s^{-1}}$.

Without a detection of the continuum flux in the spectrum of DP7,
we measured the rest-frame equivalent width (EW) for the Lyman-$\alpha$ line in the following way:
\begin{equation}
    \mathrm{EW(Ly\alpha)}= \frac{\mathrm{F_{Ly\alpha}}}{\mathrm{F_{\lambda}}(1+z)}
\end{equation}
where $\mathrm{F_{\lambda}}$ is the value of the flux density redward of Lyman-$\alpha$ from broad band photometry. We measured EW using both F105W and F140W flux densities for the entire galaxy. Because Lyman-$\alpha$ falls in the F105W band, we subtracted the Lyman-$\alpha$ flux from the broad band flux density. Moreover, we corrected the F105W flux density for the absorption blueward of Lyman-$\alpha$. When using F105W, we find that EW(Ly$\alpha$)=$237.12\pm57.78\,$\AA, while if using F140W, EW(Ly$\alpha$)=$341.55\pm93.78\,$\AA. For completeness, we also measured the EW using the monochromatic flux density from the best-fit SED template just redward of Lyman-$\alpha$ and found a consistent value ($\sim$230\AA). These values are to be considered lower limits since we did not integrate to the continuum. Moreover, we did not make a correction for the IGM absorption; therefore, the intrinsic EW is likely be to higher.

\subsection{Ly\texorpdfstring{$\alpha$}{alpha} versus F160W spatial extension} \label{subsec:lya_size}
We estimated the spatial extent of the Lyman-$\alpha$ emission by collapsing along the spectral direction a portion of the spectrum centered at the emission wavelength, and fitted a Gaussian function to the spatial profile. We determined the intrinsic FWHM of the Lyman-$\alpha$ emission, by subtracting in quadrature the seeing (see \ref{subsec:spectroscopy}), and found a value of $2.09\pm0.88\,$kpc.

As a comparison, we determined the galaxy size from the \textit{HST} F160W image. To this end, we obtained the 1D spatial profile of the galaxy along the direction of the slit (see Figure~\ref{fig:hstimage}), and determined the FWHM by fitting a Gaussian function to this profile. After accounting for the F160W point spread function, we  found that the galaxy size is $0.56\pm0.20\,$kpc. Both sizes are corrected for magnification by multiplying their values by $1/\sqrt{\mu}$.

\begin{figure}
\centering
	\includegraphics[width=\linewidth]{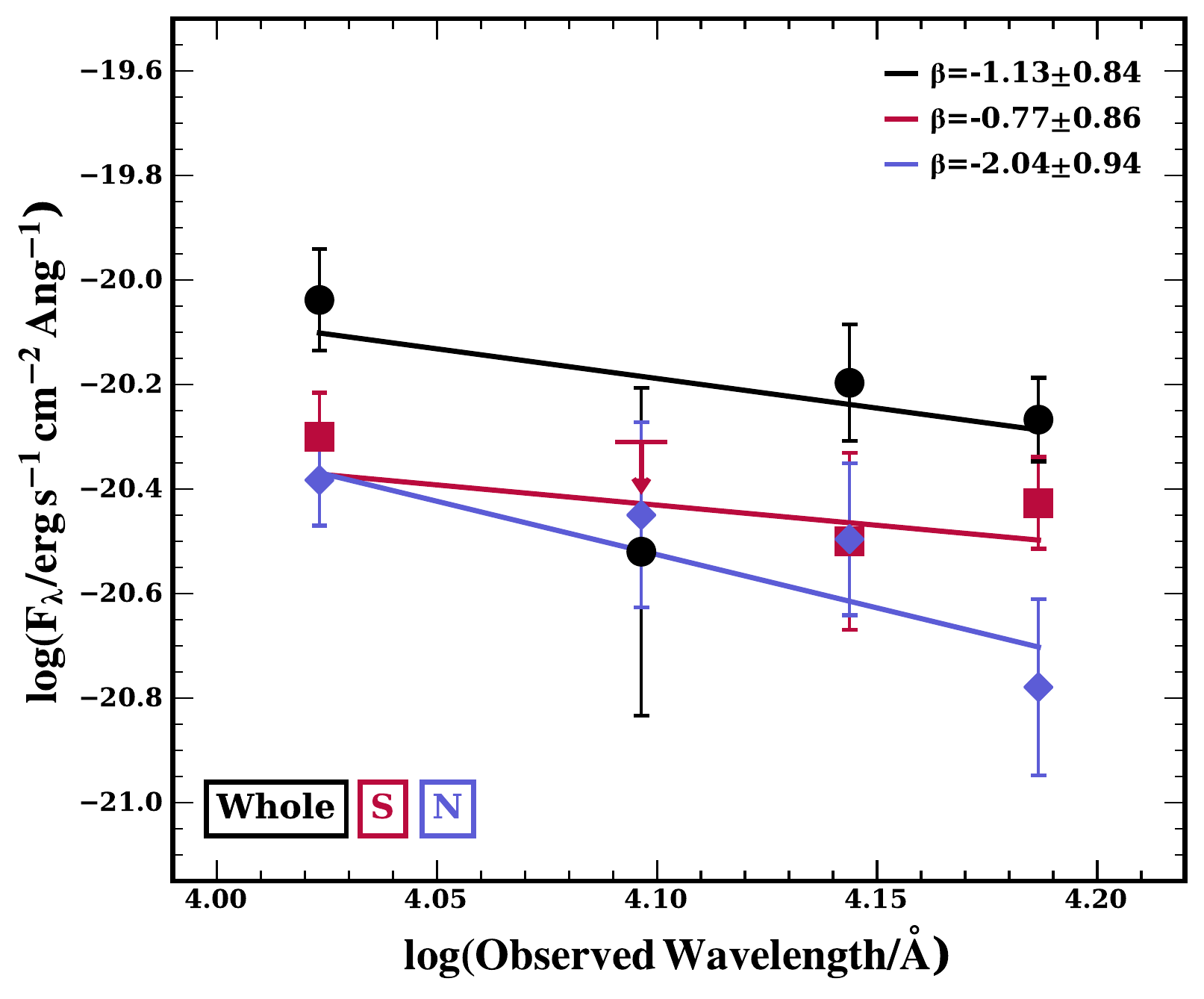}
    \caption{UV $\beta$ slope fits to photometry for the whole (black), southern (red), and northern (blue) components of DP7. }
    \label{fig:betaslope}
\end{figure}

\subsection{\texorpdfstring{UV colors and $\beta$}{beta} Slope} \label{subsec:betaslope}
From the \textit{HST} color image showed in Figure~\ref{fig:hstimage} we can clearly see that DP7 constitutes of two components: a southern red component and a northern blue component.
This is confirmed by the measured F105W-F160W colors, which are $-0.43\pm0.47$ and $0.24\pm0.30$ for the northern and southern components, respectively. Because F105W flux density is affected by the Lyman-$\alpha$ emission, we also measured F140W-F160W colors, which are $-0.49\pm0.56$ and $0.40\pm0.48$, for northern and southern components, respectively. The errors on the colors are obtained by adding in quadrature the uncertainties on the magnitudes. 

Another way of characterizing the UV colors is by investigating the slope $\beta$ of the power-law ($f_\lambda\propto\lambda^\beta$) commonly used to parametrize the galaxy's rest-frame UV continuum.
Increasing values of $\beta$ corresponds to redder galaxies. We measured $\beta$ for our whole galaxy, as well as for the two individual components identified in the \textit{HST} image, using the values of the flux densities in F105W, F125W, F140W, F160W bands. As done in Section~\ref{subsec:lya_ew} we corrected the F105W flux density from the whole galaxy for the presence of the  Lyman-$\alpha$, and assuming that the Lyman-$\alpha$ is originated from both northern and southern components, we accordingly scaled F105W flux density from both components. This assumption likely does not represent the reality, but the current Lyman-$\alpha$ spatial resolution does not allow us to determine if it originate solely from one of the two components. We fit a power-law function between flux density and wavelength, although we show it in Figure~\ref{fig:betaslope} in the logarithmic phase space for better visualization, adopting a least-squares approach that accounts for the uncertainties on the flux density. The uncertainty on $\beta$ is computed from the covariance matrix. We find that for the whole galaxy $\beta=-1.13\pm0.84$. However, when we measure the $\beta$ slope for the two individual components we find that  the northern component has $\beta=-2.04\pm0.94$, while the southern component has $\beta=-0.77\pm0.86$. Note that both for the calculation of the colors and $\beta$ slopes, we did not explicitly PSF match the various WFC3 images, though the relative difference of the components remains unchanged if we make this correction. The above $\beta$ values are consistent, though with a slightly reduced significance (1$\sigma$), with the measured color difference, implying the existence of a bluer and redder component in DP7 that have different properties in their stellar populations and/or their dust content \citep[see, e.g.,][]{Anykey2009, Fudamoto2020}.

\section{Discussion and Conclusions} \label{sec:conclusion}
We report the discovery of DP7, a Dichromatic Primeval galaxy within the RELICS survey, spectroscopically confirmed at $z=7.0281\pm0.0003$ using observations carried out with Keck/LRIS. We detect a strong Lyman-$\alpha$ emission with rest-frame EW$\,\sim300$\AA, depending on the assumed continuum. The EWs measured here are to be considered lower limits since we did not integrate to the continuum in our observations, and we did not correct for IGM absorption, which is likely to be very severe for an average galaxy at these redshifts. 

We find that Lyman-$\alpha$ is spatially extended ($2.09\pm0.88\,$kpc) and likely larger than the galaxy size ($0.56\pm0.20\,$kpc, from the \textit{HST}/F160W image), hinting to the possible existence of a Lyman-$\alpha$ halo. This phenomenon appears to be commonly seen in UV faint Lyman-$\alpha$ emitters at slightly lower redshift \citep[$z\sim2-6$, see, e.g.,][]{Wisotzki2016,Paulino-Afonso2018}. However, we cannot determine if the center of the Lyman-$\alpha$ emission coincides with the center of the UV continuum emission, due to the spatial resolution of the ground-based data used to measure the Lyman-$\alpha$ emission.

Moreover, DP7 appears dichromatic, since it is comprised of two components (see Figure~\ref{fig:hstimage}) that likely have different stellar and/or dust properties. This difference is implied by the different UV colors of the two components and, with a smaller significance, by the $\beta$ slope measurements (see Section~\ref{subsec:betaslope}), with the northern component exhibiting bluer colors ($\beta=-2.04$) and the southern component exhibits redder colors ($\beta=-0.77$).

A few recent studies have reported the spectroscopic confirmation of $z\sim7$ galaxies with properties similar to DP7, i.e, strong Lyman-$\alpha$ emission (EW$\,\gtrsim\,$100\AA), spatially extended Lyman-$\alpha$, and/or multiple components. \cite{Sobral2015} and \cite{Matthee2019} discuss MASOSA, a $z=6.5$ source, which shows many similarities with DP7: Lyman-$\alpha$ EW$\,>\,$100\AA\, an unexpectedly red $\beta$ slope ($\beta=-1.06^{+0.68}_{-0.72}$) given its UV luminosity, and a second, even redder clump. Although MASOSA has very luminous Lyman-$\alpha$, the dust continuum and [CII] $\lambda$158$\mu$m line are not detected in Atacama Large Millimeter/submillimeter Array (ALMA) observations, implying a likely young, metal-poor stellar population. \cite{Sobral2015} and \cite{Matthee2020} discuss CR7, a galaxy at $z=6.6$ with strong spatially extended Lyman-$\alpha$ halo, surrounding at least three UV components of differing color, with the redder components dominating the mass and the bluer component dominating the Lyman-$\alpha$ emission. These authors argue that the Lyman-$\alpha$ EW for both MASOSA and CR7 can be explained by a young metal-poor starburst with a possible, but not necessary, contribution from PopIII-like stellar populations. However, \cite{Vanzella2020}, who report the discovery of a highly magnified arc with intrinsic Lyman-$\alpha$ EW$\,>\,$1120\AA, argue that a PopIII stellar complex is required for emission in excess of EW(Lyman-$\alpha$)$\ga$400\AA, a value which DP7 possibly exceeds when accounting for the IGM absorption. 
Other examples of strong (although not so extreme) extended Lyman-$\alpha$ at $z\sim7$ are presented by \cite{Ouchi2013}, who discovered a Lyman-$\alpha$ halo surrounding a luminous star-forming galaxy ('Himiko') at $z=6.6$. `Himiko' is a very metal poor system and is composed by three clumps (with different colors) undergoing a triple merger, likely powering the Lyman-$\alpha$ emission. \cite{hu2016} and \cite{Matthee2018} report on the discovery of COLA1 at $z\sim6.6$. COLA1 is one of the brightest Lyman-$\alpha$ emitters at the epoch of reionization, with a Lyman-$\alpha$ luminosity of $L_{\rm{Ly}\alpha}=4.1\times10^{43}$~ergs\,s$^{-1}$ and a Lyman-$\alpha$ EW of $\sim$120\AA. These extreme properties are attributed to an extremely low gas-phase metallicity, a large ionized bubble powered by COLA1, and the possible inflow of gas. Finally, \cite{Tilvi2020} presented evidence for the existence of overlapping Lyman-$\alpha$ ionized bubbles from a grouping of three galaxies at $z=7.7$, however, no claims are made about the metallicity of these systems. 

All of these previous works suggest that extended Lyman-$\alpha$ emission with strength similar to what seen in DP7 originates from galaxies that are young and very metal poor (with a possible contribution from PopIII systems), have multiple components/companions, and/or are undergoing a merger or accretion event.
Interestingly, DP7 stands out from the UV bright galaxies with strong Lyman-$\alpha$ emission that are the subjects of the above studies in that it is one of the UV fainter galaxies spectroscopically-confirmed at these redshifts, with $L_{UV}/L^*_{UV}\sim0.3$. There are only a few other examples of galaxies at these redshifts that approach the UV faintness of DP7 and have similarly extreme Lyman-$\alpha$ \citep{Larson2018, jung20}. UV faint galaxies are typically thought to be the types of galaxies that reionized the universe (e.g, \citealt{Sawicki06}); however, recent studies (e.g., \citealp{Stark2017,Mason18b}) have suggested that reionization may be accelerated around the brightest galaxies. If systems such as DP7 are common and found without brighter companions, this could provide evidence in favor of UV faint galaxies as drivers of the reionization.

Possible scenarios for DP7 include (1) an evolved, dusty galaxy experiencing rejuvenation due to new star-forming activity and (2) a merger event between an evolved, dusty component and a younger, star-forming component. In either scenario, the strong, spatially extended Lyman-$\alpha$ emission is likely evidence of an ionized bubble explained by metal-poor star formation, with a possible contribution from PopIII stars. Investigation of the larger scale environment would allow us to look for nearby UV bright galaxies that may have carved out a bubble \citep[e.g.,][]{Tilvi2020}. Additionally, measuring the velocity offset between Lyman-$\alpha$ and systemic would allow us to probe for the presence of an ionized bubble \citep{Mason2020}. Unobscured AGN activity, instead, is ruled out due to the narrow width (see Table~\ref{tab:properties}) of Lyman-$\alpha$ emission. However, other types of AGN activity cannot be ruled out from these observations, including the presence or absence of the NV $\lambda$1240\AA\ feature (see, e.g., \citealt{Mainali2018}), which is predicted to be coincident with a sky line in the LRIS spectra. We also cannot rule out the contribution of nebular continuum (free-free) emission, which could, under extreme conditions, explain the presence of UV slopes in young star-forming galaxies that are redder than expected for a given UV luminosity \citep[see discussion in][and references therein]{Matthee2019}.

The results obtained from the observations of DP7 thus far are tantalizing but, ultimately, ambiguous. More comprehensive and deeper data are needed to understand the physical processes involved in powering such a strong Lyman-$\alpha$ emission, and to discriminate between the proposed scenarios.
Spectral observations in the near infrared to detect or constrain features such as CIV ($\lambda$1549\AA), HeII ($\lambda$1640\AA), and CIII] ($\lambda$1907,1909\AA) would help illuminate the source powering the Lyman-$\alpha$ emission including the presence of certain types of AGN activity (see \citealt{Nakajima2018, LeFevre2019}). \emph{James Webb Space Telescope} will be crucial in the study of galaxies at the epoch of reionization, in particular the IFU NIRSpec spectrograph would allow to spatially map rest-frame optical emission lines, which are very likely to be extremely strong given our lower limit on the Lyman-$\alpha$ intrinsic strength (though metallicity effects may reduce the strength of, e.g., the $\lambda$5007\AA\ [OIII] emission, see, e.g., \citealt{Matthee2018}). These observations will provide spatial distribution of star formation, ionization parameter, metallicity, and dust. Moreover, ALMA would allow us to obtain high spatial resolution observations of, e.g., [CII] emission, which can be used to better constrain the separation between the two DP7 components and investigate merger scenarios through galaxy kinematics (see, e.g., \citealt{Ginolfi2020}). Fully probing the nature of sources such as DP7 with such observations is key to understanding reionization since these types of galaxies likely represent those that make the primary contribution to the reionization of the Universe.

\acknowledgments
Support for this work was provided by NASA through NSF grant AST 1815458, NASA ADAP grant 80NSSC18K0945,  NASA \textit{HST} grant \textit{HST}-GO-14096, and through an award issued by JPL/Caltech. VS also acknowledges support through Heising-Simons Foundation
Grant \#2018-1140. Data presented here is part of the RELICS \emph{Hubble} Treasury Program (GO 14096), which consists of observations obtained by the NASA/ESA \emph{Hubble Space Telescope (\textit{HST}). Hubble} is operated by the Association of Universities for Research in Astronomy, Inc. (AURA), under NASA contract NAS5-26555. Data from the NASA/ESA \emph{Hubble Space Telescope} presented in this paper were obtained from the Mikulski Archive for Space Telescopes (MAST), operated by the Space Telescope Science Institute (STScI). STScI is operated by the Association of Universities for Research in Astronomy, Inc. (AURA) under NASA contract NAS 5-26555. The \emph{Hubble} Advanced Camera for Surveys (ACS) was developed under NASA contract NAS 5-32864. \emph{Spitzer Space Telescope} data presented in this paper were obtained from the NASA/IPAC Infrared Science Archive (IRSA), operated by the Jet Propulsion Laboratory, California Institute of Technology. Spitzer and IRSA are operated by the Jet Propulsion Laboratory, California Institute of Technology under contract with NASA.
The spectroscopic data presented herein were obtained at the W.M. Keck Observatory, which is operated as a scientific partnership among the California Institute of Technology, the University of California, and the National Aeronautics and Space Administration. The Observatory was made possible by the generous financial support of the W.M. Keck Foundation. We thank the indigenous Hawaiian community for allowing us to be guests on their sacred mountain, a privilege, without which, this work would not have been possible. We are most fortunate to be able to conduct observations from this site.

\bibliography{lymanalphaz7}{}
\bibliographystyle{aasjournal}



\end{document}

%% file: DP7_mags_no_mucorr.tex
\begin{deluxetable}{cccc}
\tabletypesize{\footnotesize}
\tablecaption{DP7 HST and Spitzer Photometry} 
\tablehead{
\colhead{\hspace{0.15cm}Filter}\hspace{0.15cm}  & \colhead{\hspace{0.15cm}Whole Galaxy}\hspace{0.15cm}  & \colhead{\hspace{0.15cm}North Component}\hspace{0.15cm} & \colhead{\hspace{0.15cm}South Component}\hspace{0.15cm}
} 
\startdata
F435W & $>28.8$         &  $>29.0$          &  $>31.5$\\
F606W & $>29.7$         &  $>30.2$          &  $>30.0$\\
F775W & $>28.7$         &  $>29.1$          &  $>29.0$\\
F105W & $27.16\pm0.15$  &  $28.02\pm0.22$   & $27.81\pm0.21$\\
F125W & $28.26\pm0.78$  &  $28.08\pm0.44$   &  $>28.9$\\
F140W & $27.21\pm0.28$  &  $27.96\pm0.36$   & $27.97\pm0.42$\\
F160W & $27.18\pm0.20$   &  $28.45\pm0.42$   &  $27.57\pm0.22$\\
$$\rm{[3.6]}$$ &  $25.4\pm0.3$  &  $--$   & $--$ \\
$$\rm{[4.5]}$$ & $25.3\pm0.4$   &  $--$   & $--$ \\
\enddata

\tablenotetext{}{We report AB magnitudes measured in HST and Spitzer images of DP7 as a whole galaxy and as two separated components, except for the Spitzer images from which we measure only magnitudes for the whole galaxy (see Section~\ref{subsec:imaging}). The magnitudes are measured in isophotal apertures shown in the inset of Figure~\ref{fig:hstimage}. }

\end{deluxetable}

%% file: properties.tex
\begin{deluxetable}{cc}
\tabletypesize{\footnotesize}
\tablecaption{\label{tbl-1} Summary of DP7 Properties} 
\tablehead{
\colhead{\hspace{1.35cm}Property}\hspace{1.35cm}  & \colhead{\hspace{1.35cm}Value}\hspace{1.35cm}
} 
\startdata
RA (Deg.) & 152.6593385 \\
Dec (Deg.) & -12.6556351 \\
$z$& $7.0281\pm0.0003$ \\
$\mu$ & $1.15\pm0.2$\\
$\rm{M}_{\rm{stellar}}$  ($10^9M_{\odot}$) & $4.9^{+3.8}_{-3.2}$\\
SFR ($M_{\odot} \rm{yr}^{-1}$) & $11.2^{+10.3}_{-7.2}$ \\ 
sSFR ($\rm{Gyr}^{-1}$) & $2.1^{+5.2}_{-2.4}$\\
E(B-V) (mag) & $0.15\pm0.10$\\
$\rm{M}_{\rm{UV}}$ (mag) &  $-19.5\pm 0.2$\\
$\mathrm{L(Ly\alpha) (erg\,s^{-1})}$ & $1.0\pm 0.1 \times 10^{43}$\\
Ly$\alpha$ EW$_{\rm{F105W}}$ (\AA) & $237.12\pm 57.78$\\
Ly$\alpha$ EW$_{\rm{F140W}}$ (\AA) & $341.55\pm 93.78$\\
Ly$\alpha$ FWHM (\kms) & $285.3^{+23.9}_{-76.4}$ \\
Ly$\alpha$ size (kpc) & $2.09\pm 0.88$\\
F160W size (kpc) & $0.56\pm 0.20$\\
\enddata

\tablenotetext{}{$\mu$ is the magnification factor: median magnification and 68\% confidence limits from the MCMC lens model uncertainties. $\mu=\mu_{\rm{med}}$ is assumed in SFR,  $\rm{M}_{\rm{stellar}}$, and $\rm{M}_{\rm{UV}}$ calculations. Uncertainties include statistical 68\% CLs from photometry and redshift. To use a different magnification value, multiply the quantity by $1/f_{\mu}$, where $f_{\mu}\equiv\mu/\mu_{\rm{med}}$. $\rm{M}_{\rm{stellar}}$ is the intrinsic stellar. sSFR is the specific SFR, sSFR $\equiv M_{stellar}$/SFR. E(B-V) is dust color excess of stellar emission. SMC dust law assumed. $\rm{M}_{\rm{UV}}$ is rest-frame UV magnitude assuming $\mu{\rm_{med}}$, derived from the observed F160W magnitude including a small template-based $k$-correction. To use a different magnification value, use $\rm{M}_{\rm{UV}}-2.5\rm{log}(f_{\mu}$). Ly$\alpha$ FWHM is measured non-parametrically by estimating the wavelength at which the line drops to 50\% of the peak on the blue and red side. The FWHM error is estimated by a Monte-Carlo technique.}

\end{deluxetable}

%% file: lymanalphaz7.bbl
\begin{thebibliography}{54}
\expandafter\ifx\csname natexlab\endcsname\relax\def\natexlab#1{#1}\fi

\bibitem[{{Anders} \& {Fritze-v.~Alvensleben}(2003)}]{Anders03}
{Anders}, P. \& {Fritze-v.~Alvensleben}, U. 2003,
  \href{http://dx.doi.org/10.1051/0004-6361:20030151}{\color{magenta}\aap},
  \href{http://adsabs.harvard.edu/abs/2003A%26A...401.1063A}{401, 1063}

\bibitem[{{Bouwens} {et~al.}(2009){Bouwens}, {Illingworth}, {Franx}, {Chary},
  {Meurer}, {Conselice}, {Ford}, {Giavalisco}, \& {van Dokkum}}]{Anykey2009}
{Bouwens}, R.~J., {Illingworth}, G.~D., {Franx}, M., {et~al.} 2009,
  \href{http://dx.doi.org/10.1088/0004-637X/705/1/936}{\color{magenta}\apj},
  \href{https://ui.adsabs.harvard.edu/abs/2009ApJ...705..936B}{705, 936}

\bibitem[{{Bouwens} {et~al.}(2015){Bouwens}, {Illingworth}, {Oesch}, {Trenti},
  {Labb{\'e}}, {Bradley}, {Carollo}, {van Dokkum}, {Gonzalez}, {Holwerda},
  {Franx}, {Spitler}, {Smit}, \& {Magee}}]{Bouwens2015a}
{Bouwens}, R.~J., {Illingworth}, G.~D., {Oesch}, P.~A., {et~al.} 2015,
  \href{http://dx.doi.org/10.1088/0004-637X/803/1/34}{\color{magenta}\apj},
  \href{https://ui.adsabs.harvard.edu/abs/2015ApJ...803...34B}{803, 34}

\bibitem[{{Brada{\v{c}}} {et~al.}(2014){Brada{\v{c}}}, {Ryan}, {Casertano},
  {Huang}, {Lemaux}, {Schrabback}, {Gonzalez}, {Allen}, {Cain}, {Gladders},
  {Hall}, {Hildebrand t}, {Hinz}, {von der Linden}, {Lubin}, {Treu}, \&
  {Zaritsky}}]{Bradac2014}
{Brada{\v{c}}}, M., {Ryan}, R., {Casertano}, S., {et~al.} 2014,
  \href{http://dx.doi.org/10.1088/0004-637X/785/2/108}{\color{magenta}\apj},
  \href{https://ui.adsabs.harvard.edu/abs/2014ApJ...785..108B}{785, 108}

\bibitem[{{Cerny} {et~al.}(2018){Cerny}, {Sharon}, {Andrade-Santos}, {Avila},
  {Brada{\v{c}}}, {Bradley}, {Carrasco}, {Coe}, {Czakon}, {Dawson}, {Frye},
  {Hoag}, {Huang}, {Johnson}, {Jones}, {Lam}, {Lovisari}, {Mainali}, {Oesch},
  {Ogaz}, {Past}, {Paterno-Mahler}, {Peterson}, {Riess}, {Rodney}, {Ryan},
  {Salmon}, {Sendra-Server}, {Stark}, {Strolger}, {Trenti}, {Umetsu},
  {Vulcani}, \& {Zitrin}}]{Cerny18}
{Cerny}, C., {Sharon}, K., {Andrade-Santos}, F., {et~al.} 2018,
  \href{http://dx.doi.org/10.3847/1538-4357/aabe7b}{\color{magenta}\apj},
  \href{https://ui.adsabs.harvard.edu/abs/2018ApJ...859..159C}{859, 159}

\bibitem[{Chabrier(2003)}]{Chabrier2003}
Chabrier, G. 2003, \href{http://dx.doi.org/10.1086/374879}{\color{magenta}The
  Astrophysical Journal}, 586, 586

\bibitem[{{Coe} {et~al.}(2019){Coe}, {Salmon}, {Brada{\v{c}}}, {Bradley},
  {Sharon}, {Zitrin}, {Acebron}, {Cerny}, {Cibirka}, {Strait},
  {Paterno-Mahler}, {Mahler}, {Avila}, {Ogaz}, {Huang}, {Pelliccia}, {Stark},
  {Mainali}, {Oesch}, {Trenti}, {Carrasco}, {Dawson}, {Rodney}, {Strolger},
  {Riess}, {Jones}, {Frye}, {Czakon}, {Umetsu}, {Vulcani}, {Graur}, {Jha},
  {Graham}, {Molino}, {Nonino}, {Hjorth}, {Selsing}, {Christensen},
  {Kikuchihara}, {Ouchi}, {Oguri}, {Welch}, {Lemaux}, {Andrade-Santos}, {Hoag},
  {Johnson}, {Peterson}, {Past}, {Fox}, {Agulli}, {Livermore}, {Ryan}, {Lam},
  {Sendra-Server}, {Toft}, {Lovisari}, \& {Su}}]{Coe2019}
{Coe}, D., {Salmon}, B., {Brada{\v{c}}}, M., {et~al.} 2019,
  \href{http://dx.doi.org/10.3847/1538-4357/ab412b}{\color{magenta}\apj},
  \href{https://ui.adsabs.harvard.edu/abs/2019ApJ...884...85C}{884, 85}

\bibitem[{{Finkelstein} {et~al.}(2013){Finkelstein}, {Papovich}, {Dickinson},
  {Song}, {Tilvi}, {Koekemoer}, {Finkelstein}, {Mobasher}, {Ferguson},
  {Giavalisco}, {Reddy}, {Ashby}, {Dekel}, {Fazio}, {Fontana}, {Grogin},
  {Huang}, {Kocevski}, {Rafelski}, {Weiner}, \& {Willner}}]{Finkelstein2013}
{Finkelstein}, S.~L., {Papovich}, C., {Dickinson}, M., {et~al.} 2013,
  \href{http://dx.doi.org/10.1038/nature12657}{\color{magenta}\nat},
  \href{https://ui.adsabs.harvard.edu/abs/2013Natur.502..524F}{502, 524}

\bibitem[{{Fudamoto} {et~al.}(2020){Fudamoto}, {Oesch}, {Faisst}, {Bethermin},
  {Ginolfi}, {Khusanova}, {Loiacono}, {Le Fevre}, {Capak}, {Schaerer},
  {Silverman}, {Cassata}, {Yan}, {Amorin}, {Bardelli}, {Boquien}, {Cimatti},
  {Dessauges-Zavadsky}, {Fujimoto}, {Gruppioni}, {Hathi}, {Ibar}, {Jones},
  {Koekemoer}, {Lagache}, {Lemaux}, {Maiolino}, {Narayanan}, {Pozzi},
  {Riechers}, {Rodighiero}, {Talia}, {Toft}, {Vallini}, {Vergani}, {Zamorani},
  \& {Zucca}}]{Fudamoto2020}
{Fudamoto}, Y., {Oesch}, P.~A., {Faisst}, A., {et~al.} 2020,
  \href{https://ui.adsabs.harvard.edu/abs/2020arXiv200410760F}{arXiv e-prints,
  arXiv:2004.10760}

\bibitem[{{Fuller} {et~al.}(2020){Fuller}, {Lemaux}, {Brada{\v{c}}}, {Hoag},
  {Schmidt}, {Huang}, {Strait}, {Mason}, {Treu}, {Pentericci}, {Trenti},
  {Henry}, \& {Malkan}}]{Fuller2020}
{Fuller}, S., {Lemaux}, B.~C., {Brada{\v{c}}}, M., {et~al.} 2020,
  \href{http://dx.doi.org/10.3847/1538-4357/ab959f}{\color{magenta}\apj},
  \href{https://ui.adsabs.harvard.edu/abs/2020ApJ...896..156F}{896, 156}

\bibitem[{{Ginolfi} {et~al.}(2020){Ginolfi}, {Jones}, {Bethermin}, {Faisst},
  {Lemaux}, {Schaerer}, {Fudamoto}, {Oesch}, {Dessauges-Zavadsky}, {Fujimoto},
  {Carniani}, {Le Fevre}, {Cassata}, {Silverman}, {Capak}, {Yan}, {Bardelli},
  {Cucciati}, {Gal}, {Gruppioni}, {Hathi}, {Lubin}, {Maiolino}, {Morselli},
  {Pelliccia}, {Talia}, {Vergani}, \& {Zamorani}}]{Ginolfi2020}
{Ginolfi}, M., {Jones}, G.~C., {Bethermin}, M., {et~al.} 2020,
  \href{https://ui.adsabs.harvard.edu/abs/2020arXiv200413737G}{arXiv e-prints,
  arXiv:2004.13737}

\bibitem[{{Hoag} {et~al.}(2019){Hoag}, {Brada{\v{c}}}, {Huang}, {Mason},
  {Treu}, {Schmidt}, {Trenti}, {Strait}, {Lemaux}, {Finney}, \&
  {Paddock}}]{hoag2019}
{Hoag}, A., {Brada{\v{c}}}, M., {Huang}, K., {et~al.} 2019,
  \href{http://dx.doi.org/10.3847/1538-4357/ab1de7}{\color{magenta}\apj},
  \href{https://ui.adsabs.harvard.edu/abs/2019ApJ...878...12H}{878, 12}

\bibitem[{{Hu} {et~al.}(2016){Hu}, {Cowie}, {Songaila}, {Barger},
  {Rosenwasser}, \& {Wold}}]{hu2016}
{Hu}, E.~M., {Cowie}, L.~L., {Songaila}, A., {et~al.} 2016,
  \href{http://dx.doi.org/10.3847/2041-8205/825/1/L7}{\color{magenta}\apjl},
  \href{https://ui.adsabs.harvard.edu/abs/2016ApJ...825L...7H}{825, L7}

\bibitem[{{Jung} {et~al.}(2020){Jung}, {Finkelstein}, {Dickinson}, {Hutchison},
  {Larson}, {Papovich}, {Pentericci}, {Straughn}, {Guo}, {Malhotra}, {Rhoads},
  {Song}, {Tilvi}, \& {Wold}}]{jung20}
{Jung}, I., {Finkelstein}, S.~L., {Dickinson}, M., {et~al.} 2020,
  \href{https://ui.adsabs.harvard.edu/abs/2020arXiv200910092J}{arXiv e-prints,
  arXiv:2009.10092}

\bibitem[{{Larson} {et~al.}(2018){Larson}, {Finkelstein}, {Pirzkal}, {Ryan},
  {Tilvi}, {Malhotra}, {Rhoads}, {Finkelstein}, {Jung}, {Christensen},
  {Cimatti}, {Ferreras}, {Grogin}, {Koekemoer}, {Hathi}, {O'Connell},
  {{\"O}stlin}, {Pasquali}, {Pharo}, {Rothberg}, {Windhorst}, \& {FIGS
  Team}}]{Larson2018}
{Larson}, R.~L., {Finkelstein}, S.~L., {Pirzkal}, N., {et~al.} 2018,
  \href{http://dx.doi.org/10.3847/1538-4357/aab893}{\color{magenta}\apj},
  \href{https://ui.adsabs.harvard.edu/abs/2018ApJ...858...94L}{858, 94}

\bibitem[{{Le F{\`e}vre} {et~al.}(2019){Le F{\`e}vre}, {Lemaux}, {Nakajima},
  {Schaerer}, {Talia}, {Zamorani}, {Cassata}, {Garilli}, {Maccagni},
  {Pentericci}, {Tasca}, {Zucca}, {Amorin}, {Bardelli}, {Cimatti},
  {Giavalisco}, {Guaita}, {Hathi}, {Marchi}, {Vanzella}, {Vergani}, \&
  {Dunlop}}]{LeFevre2019}
{Le F{\`e}vre}, O., {Lemaux}, B.~C., {Nakajima}, K., {et~al.} 2019,
  \href{http://dx.doi.org/10.1051/0004-6361/201732197}{\color{magenta}\aap},
  \href{https://ui.adsabs.harvard.edu/abs/2019A&A...625A..51L}{625, A51}

\bibitem[{{Leitherer} \& {Heckman}(1995)}]{Leitherer95}
{Leitherer}, C. \& {Heckman}, T.~M. 1995,
  \href{http://dx.doi.org/10.1086/192112}{\color{magenta}\apjs},
  \href{http://adsabs.harvard.edu/abs/1995ApJS...96....9L}{96, 9}

\bibitem[{{Lemaux} {et~al.}(2009){Lemaux}, {Lubin}, {Sawicki}, {Martin},
  {Lagattuta}, {Gal}, {Kocevski}, {Fassnacht}, \& {Squires}}]{Lemaux2009}
{Lemaux}, B.~C., {Lubin}, L.~M., {Sawicki}, M., {et~al.} 2009,
  \href{http://dx.doi.org/10.1088/0004-637X/700/1/20}{\color{magenta}\apj},
  \href{https://ui.adsabs.harvard.edu/abs/2009ApJ...700...20L}{700, 20}

\bibitem[{{Mainali} {et~al.}(2018){Mainali}, {Zitrin}, {Stark}, {Ellis},
  {Richard}, {Tang}, {Laporte}, {Oesch}, \& {McGreer}}]{Mainali2018}
{Mainali}, R., {Zitrin}, A., {Stark}, D.~P., {et~al.} 2018,
  \href{http://dx.doi.org/10.1093/mnras/sty1640}{\color{magenta}\mnras},
  \href{https://ui.adsabs.harvard.edu/abs/2018MNRAS.479.1180M}{479, 1180}

\bibitem[{{Malhotra} \& {Rhoads}(2006)}]{malhotra06}
{Malhotra}, S. \& {Rhoads}, J.~E. 2006,
  \href{http://dx.doi.org/10.1086/506983}{\color{magenta}\apjl},
  \href{https://ui.adsabs.harvard.edu/abs/2006ApJ...647L..95M}{647, L95}

\bibitem[{{Mason} \& {Gronke}(2020)}]{Mason2020}
{Mason}, C.~A. \& {Gronke}, M. 2020,
  \href{http://dx.doi.org/10.1093/mnras/staa2910}{\color{magenta}\mnras},
  \href{https://ui.adsabs.harvard.edu/abs/2020MNRAS.499.1395M}{499, 1395}

\bibitem[{{Mason} {et~al.}(2019){Mason}, {Naidu}, {Tacchella}, \&
  {Leja}}]{Mason2019}
{Mason}, C.~A., {Naidu}, R.~P., {Tacchella}, S., \& {Leja}, J. 2019,
  \href{http://dx.doi.org/10.1093/mnras/stz2291}{\color{magenta}\mnras},
  \href{https://ui.adsabs.harvard.edu/abs/2019MNRAS.489.2669M}{489, 2669}

\bibitem[{{Mason} {et~al.}(2018{\natexlab{a}}){Mason}, {Treu}, {de Barros},
  {Dijkstra}, {Fontana}, {Mesinger}, {Pentericci}, {Trenti}, \&
  {Vanzella}}]{Mason18b}
{Mason}, C.~A., {Treu}, T., {de Barros}, S., {et~al.} 2018{\natexlab{a}},
  \href{http://dx.doi.org/10.3847/2041-8213/aabbab}{\color{magenta}\apjl},
  \href{https://ui.adsabs.harvard.edu/abs/2018ApJ...857L..11M}{857, L11}

\bibitem[{{Mason} {et~al.}(2018{\natexlab{b}}){Mason}, {Treu}, {Dijkstra},
  {Mesinger}, {Trenti}, {Pentericci}, {de Barros}, \& {Vanzella}}]{mason18a}
{Mason}, C.~A., {Treu}, T., {Dijkstra}, M., {et~al.} 2018{\natexlab{b}},
  \href{http://dx.doi.org/10.3847/1538-4357/aab0a7}{\color{magenta}\apj},
  \href{http://adsabs.harvard.edu/abs/2018ApJ...856....2M}{856, 2}

\bibitem[{{Matthee} {et~al.}(2020){Matthee}, {Pezzulli}, {Mackenzie},
  {Cantalupo}, {Kusakabe}, {Leclercq}, {Sobral}, {Richard}, {Wisotzki},
  {Lilly}, {Boogaard}, {Marino}, {Maseda}, \& {Nanayakkara}}]{Matthee2020}
{Matthee}, J., {Pezzulli}, G., {Mackenzie}, R., {et~al.} 2020,
  \href{http://dx.doi.org/10.1093/mnras/staa2550}{\color{magenta}\mnras},
  \href{https://ui.adsabs.harvard.edu/abs/2020MNRAS.498.3043M}{498, 3043}

\bibitem[{{Matthee} {et~al.}(2019){Matthee}, {Sobral}, {Boogaard},
  {R{\"o}ttgering}, {Vallini}, {Ferrara}, {Paulino-Afonso}, {Boone},
  {Schaerer}, \& {Mobasher}}]{Matthee2019}
{Matthee}, J., {Sobral}, D., {Boogaard}, L.~A., {et~al.} 2019,
  \href{http://dx.doi.org/10.3847/1538-4357/ab2f81}{\color{magenta}\apj},
  \href{https://ui.adsabs.harvard.edu/abs/2019ApJ...881..124M}{881, 124}

\bibitem[{{Matthee} {et~al.}(2017){Matthee}, {Sobral}, {Boone},
  {R{\"o}ttgering}, {Schaerer}, {Girard}, {Pallottini}, {Vallini}, {Ferrara},
  {Darvish}, \& {Mobasher}}]{Matthee2017b}
{Matthee}, J., {Sobral}, D., {Boone}, F., {et~al.} 2017,
  \href{http://dx.doi.org/10.3847/1538-4357/aa9931}{\color{magenta}\apj},
  \href{https://ui.adsabs.harvard.edu/abs/2017ApJ...851..145M}{851, 145}

\bibitem[{{Matthee} {et~al.}(2018){Matthee}, {Sobral}, {Gronke},
  {Paulino-Afonso}, {Stefanon}, \& {R{\"o}ttgering}}]{Matthee2018}
{Matthee}, J., {Sobral}, D., {Gronke}, M., {et~al.} 2018,
  \href{http://dx.doi.org/10.1051/0004-6361/201833528}{\color{magenta}\aap},
  \href{https://ui.adsabs.harvard.edu/abs/2018A&A...619A.136M}{619, A136}

\bibitem[{{Nakajima} {et~al.}(2018){Nakajima}, {Schaerer}, {Le F{\`e}vre},
  {Amor{\'\i}n}, {Talia}, {Lemaux}, {Tasca}, {Vanzella}, {Zamorani},
  {Bardelli}, {Grazian}, {Guaita}, {Hathi}, {Pentericci}, \&
  {Zucca}}]{Nakajima2018}
{Nakajima}, K., {Schaerer}, D., {Le F{\`e}vre}, O., {et~al.} 2018,
  \href{http://dx.doi.org/10.1051/0004-6361/201731935}{\color{magenta}\aap},
  \href{https://ui.adsabs.harvard.edu/abs/2018A&A...612A..94N}{612, A94}

\bibitem[{{Oesch} {et~al.}(2015){Oesch}, {van Dokkum}, {Illingworth},
  {Bouwens}, {Momcheva}, {Holden}, {Roberts-Borsani}, {Smit}, {Franx},
  {Labb{\'e}}, {Gonz{\'a}lez}, \& {Magee}}]{Oesch2015}
{Oesch}, P.~A., {van Dokkum}, P.~G., {Illingworth}, G.~D., {et~al.} 2015,
  \href{http://dx.doi.org/10.1088/2041-8205/804/2/L30}{\color{magenta}\apjl},
  \href{https://ui.adsabs.harvard.edu/abs/2015ApJ...804L..30O}{804, L30}

\bibitem[{{Ouchi} {et~al.}(2013){Ouchi}, {Ellis}, {Ono}, {Nakanishi}, {Kohno},
  {Momose}, {Kurono}, {Ashby}, {Shimasaku}, {Willner}, {Fazio}, {Tamura}, \&
  {Iono}}]{Ouchi2013}
{Ouchi}, M., {Ellis}, R., {Ono}, Y., {et~al.} 2013,
  \href{http://dx.doi.org/10.1088/0004-637X/778/2/102}{\color{magenta}\apj},
  \href{https://ui.adsabs.harvard.edu/abs/2013ApJ...778..102O}{778, 102}

\bibitem[{{Paulino-Afonso} {et~al.}(2018){Paulino-Afonso}, {Sobral}, {Ribeiro},
  {Matthee}, {Santos}, {Calhau}, {Forshaw}, {Johnson}, {Merrick}, {P{\'e}rez},
  \& {Sheldon}}]{Paulino-Afonso2018}
{Paulino-Afonso}, A., {Sobral}, D., {Ribeiro}, B., {et~al.} 2018,
  \href{http://dx.doi.org/10.1093/mnras/sty281}{\color{magenta}\mnras},
  \href{https://ui.adsabs.harvard.edu/abs/2018MNRAS.476.5479P}{476, 5479}

\bibitem[{{Pentericci} {et~al.}(2014){Pentericci}, {Vanzella}, {Fontana},
  {Castellano}, {Treu}, {Mesinger}, {Dijkstra}, {Grazian}, {Brada{\v c}},
  {Conselice}, {Cristiani}, {Dunlop}, {Galametz}, {Giavalisco}, {Giallongo},
  {Koekemoer}, {McLure}, {Maiolino}, {Paris}, \& {Santini}}]{pentericci14}
{Pentericci}, L., {Vanzella}, E., {Fontana}, A., {et~al.} 2014,
  \href{http://dx.doi.org/10.1088/0004-637X/793/2/113}{\color{magenta}\apj},
  \href{http://adsabs.harvard.edu/abs/2014ApJ...793..113P}{793, 113}

\bibitem[{{Prochaska} {et~al.}(2020){Prochaska}, {Hennawi}, {Westfall},
  {Cooke}, {Wang}, {Hsyu}, {Davies}, \& {Farina}}]{Prochaska2020}
{Prochaska}, J.~X., {Hennawi}, J.~F., {Westfall}, K.~B., {et~al.} 2020,
  \href{https://ui.adsabs.harvard.edu/abs/2020arXiv200506505P}{arXiv e-prints,
  arXiv:2005.06505}

\bibitem[{{Roberts-Borsani} {et~al.}(2016){Roberts-Borsani}, {Bouwens},
  {Oesch}, {Labbe}, {Smit}, {Illingworth}, {van Dokkum}, {Holden}, {Gonzalez},
  {Stefanon}, {Holwerda}, \& {Wilkins}}]{Roberts-Borsani2016}
{Roberts-Borsani}, G.~W., {Bouwens}, R.~J., {Oesch}, P.~A., {et~al.} 2016,
  \href{http://dx.doi.org/10.3847/0004-637X/823/2/143}{\color{magenta}\apj},
  \href{https://ui.adsabs.harvard.edu/abs/2016ApJ...823..143R}{823, 143}

\bibitem[{{Robertson} {et~al.}(2015){Robertson}, {Ellis}, {Furlanetto}, \&
  {Dunlop}}]{robertson15}
{Robertson}, B.~E., {Ellis}, R.~S., {Furlanetto}, S.~R., \& {Dunlop}, J.~S.
  2015,
  \href{http://dx.doi.org/10.1088/2041-8205/802/2/L19}{\color{magenta}\apjl},
  \href{http://adsabs.harvard.edu/abs/2015ApJ...802L..19R}{802, L19}

\bibitem[{{Salmon} {et~al.}(2020){Salmon}, {Coe}, {Bradley}, {Bouwens},
  {Brada{\v{c}}}, {Huang}, {Oesch}, {Stark}, {Sharon}, {Trenti}, {Avila},
  {Ogaz}, {Andrade-Santos}, {Carrasco}, {Cerny}, {Dawson}, {Frye}, {Hoag},
  {Johnson}, {Jones}, {Lam}, {Lovisari}, {Mainali}, {Past}, {Paterno-Mahler},
  {Peterson}, {Riess}, {Rodney}, {Ryan}, {Sendra-Server}, {Strait}, {Strolger},
  {Umetsu}, {Vulcani}, \& {Zitrin}}]{Salmon2020}
{Salmon}, B., {Coe}, D., {Bradley}, L., {et~al.} 2020,
  \href{http://dx.doi.org/10.3847/1538-4357/ab5a8b}{\color{magenta}\apj},
  \href{https://ui.adsabs.harvard.edu/abs/2020ApJ...889..189S}{889, 189}

\bibitem[{{Sawicki} \& {Thompson}(2006)}]{Sawicki06}
{Sawicki}, M. \& {Thompson}, D. 2006,
  \href{http://dx.doi.org/10.1086/505902}{\color{magenta}\apj},
  \href{https://ui.adsabs.harvard.edu/abs/2006ApJ...648..299S}{648, 299}

\bibitem[{{Schenker} {et~al.}(2014){Schenker}, {Ellis}, {Konidaris}, \&
  {Stark}}]{schenker14}
{Schenker}, M.~A., {Ellis}, R.~S., {Konidaris}, N.~P., \& {Stark}, D.~P. 2014,
  \href{http://dx.doi.org/10.1088/0004-637X/795/1/20}{\color{magenta}\apj},
  \href{http://adsabs.harvard.edu/abs/2014ApJ...795...20S}{795, 20}

\bibitem[{{Sharon} {et~al.}(2020){Sharon}, {Bayliss}, {Dahle}, {Dunham},
  {Florian}, {Gladders}, {Johnson}, {Mahler}, {Paterno-Mahler}, {Rigby},
  {Whitaker}, {Akhshik}, {Koester}, {Murray}, {Remolina Gonz{\'a}lez}, \&
  {Wuyts}}]{Sharon20}
{Sharon}, K., {Bayliss}, M.~B., {Dahle}, H., {et~al.} 2020,
  \href{http://dx.doi.org/10.3847/1538-4365/ab5f13}{\color{magenta}\apjs},
  \href{https://ui.adsabs.harvard.edu/abs/2020ApJS..247...12S}{247, 12}

\bibitem[{{Smit} {et~al.}(2014){Smit}, {Bouwens}, {Labb{\'e}}, {Zheng},
  {Bradley}, {Donahue}, {Lemze}, {Moustakas}, {Umetsu}, {Zitrin}, {Coe},
  {Postman}, {Gonzalez}, {Bartelmann}, {Ben{\'{\i}}tez}, {Broadhurst}, {Ford},
  {Grillo}, {Infante}, {Jimenez-Teja}, {Jouvel}, {Kelson}, {Lahav}, {Maoz},
  {Medezinski}, {Melchior}, {Meneghetti}, {Merten}, {Molino}, {Moustakas},
  {Nonino}, {Rosati}, \& {Seitz}}]{Smit14}
{Smit}, R., {Bouwens}, R.~J., {Labb{\'e}}, I., {et~al.} 2014,
  \href{http://dx.doi.org/10.1088/0004-637X/784/1/58}{\color{magenta}\apj},
  \href{http://adsabs.harvard.edu/abs/2014ApJ...784...58S}{784, 58}

\bibitem[{{Sobral} \& {Matthee}(2019)}]{Sobral_Matthee2019}
{Sobral}, D. \& {Matthee}, J. 2019,
  \href{http://dx.doi.org/10.1051/0004-6361/201833075}{\color{magenta}\aap},
  \href{https://ui.adsabs.harvard.edu/abs/2019A&A...623A.157S}{623, A157}

\bibitem[{{Sobral} {et~al.}(2015){Sobral}, {Matthee}, {Darvish}, {Schaerer},
  {Mobasher}, {R{\"o}ttgering}, {Santos}, \& {Hemmati}}]{Sobral2015}
{Sobral}, D., {Matthee}, J., {Darvish}, B., {et~al.} 2015,
  \href{http://dx.doi.org/10.1088/0004-637X/808/2/139}{\color{magenta}\apj},
  \href{https://ui.adsabs.harvard.edu/abs/2015ApJ...808..139S}{808, 139}

\bibitem[{{Stark} {et~al.}(2017){Stark}, {Ellis}, {Charlot}, {Chevallard},
  {Tang}, {Belli}, {Zitrin}, {Mainali}, {Gutkin}, {Vidal-Garc{\'\i}a},
  {Bouwens}, \& {Oesch}}]{Stark2017}
{Stark}, D.~P., {Ellis}, R.~S., {Charlot}, S., {et~al.} 2017,
  \href{http://dx.doi.org/10.1093/mnras/stw2233}{\color{magenta}\mnras},
  \href{https://ui.adsabs.harvard.edu/abs/2017MNRAS.464..469S}{464, 469}

\bibitem[{{Stark} {et~al.}(2010){Stark}, {Ellis}, {Chiu}, {Ouchi}, \&
  {Bunker}}]{stark10}
{Stark}, D.~P., {Ellis}, R.~S., {Chiu}, K., {Ouchi}, M., \& {Bunker}, A. 2010,
  \href{http://dx.doi.org/10.1111/j.1365-2966.2010.17227.x}{\color{magenta}\mnras},
  \href{http://adsabs.harvard.edu/abs/2010MNRAS.408.1628S}{408, 1628}

\bibitem[{{Strait} {et~al.}(2020){Strait}, {Bradac}, {Coe}, {Lemaux},
  {Carnall}, {Bradley}, {Pelliccia}, {Sharon}, {Zitrin}, {Acebron},
  {Andrade-Santos}, {Avila}, {Frye}, {Mahler}, {Nonino}, {Ogaz}, {Oguri},
  {Ouchi}, {Paterno-Mahler}, {Stark}, {Mainali}, {Oesch}, {Trenti}, {Carrasco},
  {Dawson}, {Jones}, {Umetsu}, \& {Vulcani}}]{Strait2020b}
{Strait}, V., {Bradac}, M., {Coe}, D., {et~al.} 2020,
  \href{https://ui.adsabs.harvard.edu/abs/2020arXiv200900020S}{arXiv e-prints,
  arXiv:2009.00020}

\bibitem[{{Tachibana} \& {Miller}(2018)}]{TachibanaMiller2018}
{Tachibana}, Y. \& {Miller}, A.~A. 2018,
  \href{http://dx.doi.org/10.1088/1538-3873/aae3d9}{\color{magenta}\pasp},
  \href{https://ui.adsabs.harvard.edu/abs/2018PASP..130l8001T}{130, 128001}

\bibitem[{{Tilvi} {et~al.}(2020){Tilvi}, {Malhotra}, {Rhoads}, {Coughlin},
  {Zheng}, {Finkelstein}, {Veilleux}, {Mobasher}, {Wang}, {Probst}, {Swaters},
  {Hibon}, {Joshi}, {Zabl}, {Jiang}, {Pharo}, \& {Yang}}]{Tilvi2020}
{Tilvi}, V., {Malhotra}, S., {Rhoads}, J.~E., {et~al.} 2020,
  \href{http://dx.doi.org/10.3847/2041-8213/ab75ec}{\color{magenta}\apjl},
  \href{https://ui.adsabs.harvard.edu/abs/2020ApJ...891L..10T}{891, L10}

\bibitem[{{Trenti} {et~al.}(2011){Trenti}, {Bradley}, {Stiavelli}, {Oesch},
  {Treu}, {Bouwens}, {Shull}, {MacKenty}, {Carollo}, \&
  {Illingworth}}]{Trenti2011}
{Trenti}, M., {Bradley}, L.~D., {Stiavelli}, M., {et~al.} 2011,
  \href{http://dx.doi.org/10.1088/2041-8205/727/2/L39}{\color{magenta}\apjl},
  \href{https://ui.adsabs.harvard.edu/abs/2011ApJ...727L..39T}{727, L39}

\bibitem[{{Treu} {et~al.}(2013){Treu}, {Schmidt}, {Trenti}, {Bradley}, \&
  {Stiavelli}}]{treu13}
{Treu}, T., {Schmidt}, K.~B., {Trenti}, M., {Bradley}, L.~D., \& {Stiavelli},
  M. 2013,
  \href{http://dx.doi.org/10.1088/2041-8205/775/1/L29}{\color{magenta}\apjl},
  \href{http://adsabs.harvard.edu/abs/2013ApJ...775L..29T}{775, L29}

\bibitem[{{Vanzella} {et~al.}(2020){Vanzella}, {Meneghetti}, {Caminha},
  {Castellano}, {Calura}, {Rosati}, {Grillo}, {Dijkstra}, {Gronke}, {Sani},
  {Mercurio}, {Tozzi}, {Nonino}, {Cristiani}, {Mignoli}, {Pentericci}, {Gilli},
  {Treu}, {Caputi}, {Cupani}, {Fontana}, {Grazian}, \&
  {Balestra}}]{Vanzella2020}
{Vanzella}, E., {Meneghetti}, M., {Caminha}, G.~B., {et~al.} 2020,
  \href{http://dx.doi.org/10.1093/mnrasl/slaa041}{\color{magenta}\mnras},
  \href{https://ui.adsabs.harvard.edu/abs/2020MNRAS.494L..81V}{494, L81}

\bibitem[{{Verhamme} {et~al.}(2015){Verhamme}, {Orlitov{\'a}}, {Schaerer}, \&
  {Hayes}}]{Verhamme2015}
{Verhamme}, A., {Orlitov{\'a}}, I., {Schaerer}, D., \& {Hayes}, M. 2015,
  \href{http://dx.doi.org/10.1051/0004-6361/201423978}{\color{magenta}\aap},
  \href{https://ui.adsabs.harvard.edu/abs/2015A&A...578A...7V}{578, A7}

\bibitem[{{Wisotzki} {et~al.}(2016){Wisotzki}, {Bacon}, {Blaizot},
  {Brinchmann}, {Herenz}, {Schaye}, {Bouch{\'e}}, {Cantalupo}, {Contini},
  {Carollo}, {Caruana}, {Courbot}, {Emsellem}, {Kamann}, {Kerutt}, {Leclercq},
  {Lilly}, {Patr{\'\i}cio}, {Sandin}, {Steinmetz}, {Straka}, {Urrutia},
  {Verhamme}, {Weilbacher}, \& {Wendt}}]{Wisotzki2016}
{Wisotzki}, L., {Bacon}, R., {Blaizot}, J., {et~al.} 2016,
  \href{http://dx.doi.org/10.1051/0004-6361/201527384}{\color{magenta}\aap},
  \href{https://ui.adsabs.harvard.edu/abs/2016A&A...587A..98W}{587, A98}

\bibitem[{{Zitrin} {et~al.}(2015){Zitrin}, {Labb{\'e}}, {Belli}, {Bouwens},
  {Ellis}, {Roberts-Borsani}, {Stark}, {Oesch}, \& {Smit}}]{Zitrin2015}
{Zitrin}, A., {Labb{\'e}}, I., {Belli}, S., {et~al.} 2015,
  \href{http://dx.doi.org/10.1088/2041-8205/810/1/L12}{\color{magenta}\apjl},
  \href{https://ui.adsabs.harvard.edu/abs/2015ApJ...810L..12Z}{810, L12}

\end{thebibliography}
